# Twitter as a Transport Layer Platform

Dmitry Namiot
Lomonosov Moscow State University
Moscow, Russia
dnamiot@gmail.com

*Abstract*—Internet messengers and social networks have become an integral part of modern digital life. We have in mind not only the interaction between individual users but also a variety of applications that exist in these applications. Typically, applications for social networks use the universal login system and rely on data from social networks. Also, such applications are likely to get more traction when they are inside of the big social network like Facebook. At the same time, less attention is paid to communication capabilities of social networks. In this paper, we target Twitter as a messaging system at the first hand. We describe the way information systems can use Twitter as a transport layer for own services. Our work introduces a programmable service called 411 for Twitter, which supports user-defined and application-specific commands through tweets.

I. INTRODUCTION

All social networks nowadays offer some public Application Program Interfaces (API). All social networks offer the ability for third-party developers to build applications. APIs enable developers to reuse the basic functionality of social networks in own projects. For example, public API for Twitter lets developers (third party applications) post new tweets, search for tweets, etc. In other words, developers can "embed" some part of a functionality of Twitter's client into own code.

But the main idea for the most of "social-connected" applications is still either share data in social networks or collect shared data. All actions are performed with the hope of the viral effect of social media. If we can make our content engaging, interesting and important enough, people will spread it widely and our content will have a disproportionate impact.

For example, our own application redefined geo-check-in (a status message with geo-coordinates) with an idea to attract more users from the huge user base of Facebook [1].

In this paper, we would like to discuss another aspect of social networks software. Social media besides data sharing (or more precisely – for data sharing support) should maintain the connectivity between own users. Social media engagement is a core part of any social media strategy. Users (especially, mobile users) spend more and more time in social media. As per [2], 94% of first-year college students use social networking websites. These data are congruent with more recent statistics on social networking website use and reinforce the fact that social networking is an important part of college students' lives. In our paper, we target Twitter. Twitter is a real-time information network where people can discover what's happening in the world right now, share information instantly and connect with people and businesses around the globe [3]. Twitter has over 300 million monthly active users. As per official statistics, 500 million tweets (messages) sent every day. The company states that 80% of users on Twitter are accessing it via a mobile device. As per official page, the company suggests the following business areas for Twitter:

Business can find out what's going on in the own industry and what the customers are interested in. Business can use Twitter search to listen to the relevant conversations that are happening and jump in where a value could be added.

Twitter has got a flat structure for social circles. It is very easy to connect with anyone. So, it is easy to start new discussions (and/or join existing discussions).

Business can raise the profile and increase the impact of own marketing efforts by using Twitter to regularly communicate with own customers. For example, Twitter suggests extending the reach even further with Twitter Ads.

The last point is especially interested in working for us. Twitter (as a company) suggests the usage of the system for providing customer service. As per many polls, more than 70% of Twitter users said Twitter provides them with a quick way to reply to customer service issues [4]. Business can use Twitter to quickly and easily respond to support queries. In an education, it could give students a low-stress way to ask questions. As it is mentioned in [5], first-year and/or introverted students are less comfortable asking questions in class. The dynamics of Twitter allow students to feel more comfortable asking questions given the psychological barriers inherent in online communication. Twitter could be used for providing academic and personal support. It could be used for delivery information about academic enrichment opportunities on campus (for instance, the location and hours for the tutoring center) in response to student requests for help [5].

Actually, there are no requirements to the "manual" only responses. Why do not allow applications respond to queries? It is the main topic for our paper. The core idea is very transparent. All social media in addition to the exchange of data (to be exact - for its support) should provide some form of a link between its users. If users are spending more and more time on social networks, is it possible to use the connecting mechanisms of the social media for delivering to users in social media data from other applications? It is not about programming for social networks. It is about data delivery via social networks. It is about the embedding data transfer mechanisms from social networks (social media) into existing applications.

The rest of the paper is organized as follows. In Section II, we briefly describe Twitter API. In Section III, we discuss the related projects. And Section IV is devoted to our Twitter 411 approach.

## II. TWITTER API

We have mentioned Twitter due to several reasons. Historically, public API (Application Program Interface) for Twitter was one of the most popular social APIs across social web developers. Twitter is a social media and a network messenger in the same time. Since its launch in 2006, Twitter has become one of the most important social properties on the web. Actually, Twitter promoted the growth and engagement of third party websites through its API.

There are two main offerings in Twitter API interested for our tasks: REST API and Streaming API.

The REST APIs provides programmatic access to read and write Twitter data. It is possible, for example, to publish a new Tweet, read Tweets, read author's profile, etc. The REST API identifies Twitter applications and users using OAuth; responses are available in JSON [6].

The Streaming APIs continuously deliver new responses to REST API queries over a long-lived HTTP connection. It lets receive updates on the latest Tweets matching a search query, stay in sync with user profile updates, etc.

Connecting to the streaming API requires keeping a persistent HTTP connection open. The difference from REST API is illustrated in figures 1 and 2 [7].

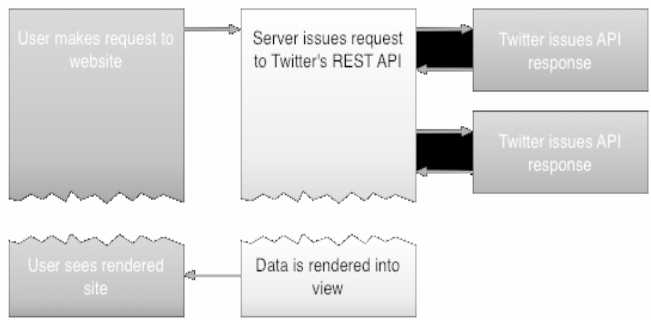

Fig. 1. Twitter REST API model [7]

In REST API an application (e.g., a web application) accepts user requests, makes one or more requests to Twitter's API, then formats and prints the result to the user, as a response to the user's initial request.

An application which connects to the Streaming APIs will not be able to establish a connection in response to a user request. Instead, the code for maintaining the Streaming connection is typically run in a process separate from the process which handles HTTP requests. In Figure 2 we have two server processes, where one process receives streamed Tweets, while the other handles HTTP requests.

Twitter's APIs are subject to rate limits. Streaming API limits and REST API limits are completely separate entities. So, obtaining entities via a streaming API doesn't consume any REST API rate limits.

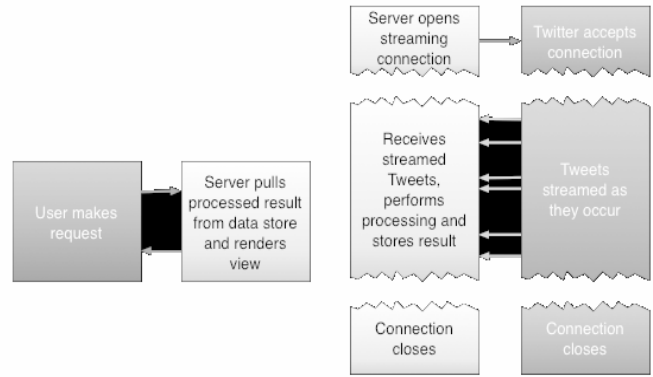

Fig. 2. Twitter Streaming API [7]

In our project, we propose the customized replies to messages (statuses) in Twitter for any selected account. Technically, there are two ways to "address" a message in Twitter to the particular account. At the first hand, it is so-called "mention". In this case, the status just contains a name for the targeted account. And the second way is so-called direct messages. In this case, the status (message) has got an immediate recipient. The main difference is the visibility. The mentions (replies) are potentially visible for other readers, the direct messages are private.

## III. RELATED WORKS

As a basic prototype model for our service, we used the well-known scheme of a functioning of information services based on Short Message Service (SMS).

How does this model work? There is a certain service number, where incoming messages could be processed programmatically (by the special software). The pair "service number" describes here the assigned functionality only. Technically, this is an ordinary telephone number to which you can send SMS. In the simplest case, for this kind of system, we can use a regular cell phone and a so-called data cable. The phone can be connected with the computer and receive from him the standard AT- command. With these commands, we can read incoming SMS, programmatically process them and send the answers (also with the help of AT-commands) [8].

The whole model is the classic question-answer system. Users send queries via SMS text and receive answers via SMS too. If the text of the response exceeds 140 characters, it can be designed as a web page. The link to that page (as an original or short URL) could be sent as a response. All existing SMS-customers on mobile phones detect links in emails and let you open them directly from the text.

The issue of payment for such access to information is, of course, not a primary question. Technically, this scheme of SMS processing does not depend on charging. Usually, this kind of services are designed for specific tasks (information systems) [9,10]. At the same time, there are a kind of toolbox (development tool system), which allow to design this kind of services [11].

In general, such a model can be described as the deployment of SMS as the transport layer in information systems. Of course, instead of SMS, we can use multimedia messaging (MMS) too [12]. Up to this time, the information systems based on SMS are an important channel to deliver content to mobile users. Actually, it is not least due to the convenience of receiving the payments for the delivered content.

SMS can be used in payment systems [13,14] too. In fact - it's the same usage of SMS as a transport layer. Our own services use SMS as transport for delivering geo-location data [15,16].

All the above-mentioned systems have a common feature – they use a part of the existing service (the transportation component) in their applications. In the above cases, it was a part of the service of telecommunications operators. But currently we see an obvious trend in switching to Internet services from the pure telecom offerings [17]. In our particular case, we want to "borrow" the transportation component of the social networks (Twitter).

We can mention in this connection so-called Tweet-a-Program application from Wolfram Alpha [18]. It lets you compose a tweet-length Wolfram Language program and tweet it to @WolframTaP account. Their Twitter bot will run your program in the Wolfram Cloud and tweet back the result. It is illustrated in Figure 3:

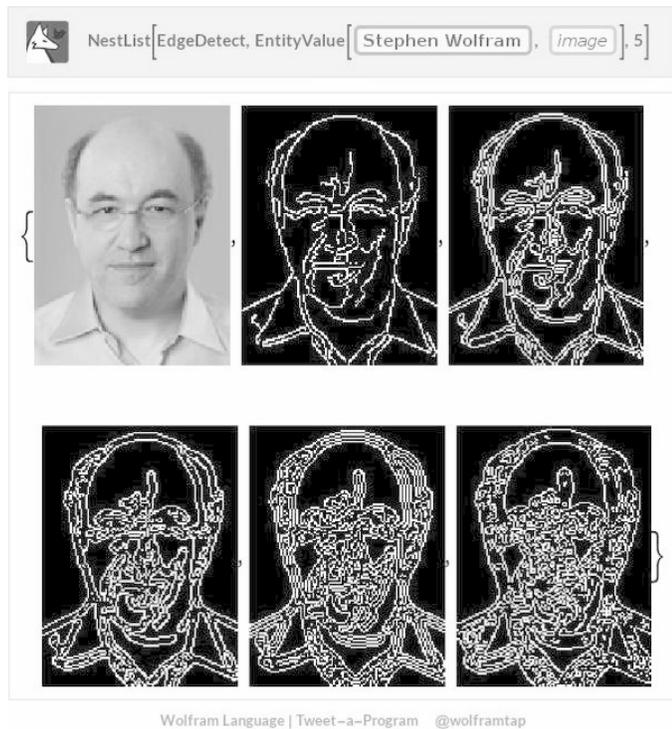

Fig. 3. Tweet-a-Program and its response [18].

Actually, such an "one line mathematics" could be quite powerful [19]. An "one-tweet" (140 characters) application in Figure 4 is 140-character expression that produces a graph that shows which countries (indicated by their flags) share borders.

XMPP protocol and tools, based on this protocol have a long history of data delivery automation (chat bots, for example) [20].

Authors in [21] present a tool for chat bots programming. Chat bot (in their interpretation) is an educational software tool whose design goal is to motivate students to learn basic Computer Science concepts through the construction of automated chats. They should be programmed to answer in different ways depending on who it is talking to, previous replies, talked topics, etc.

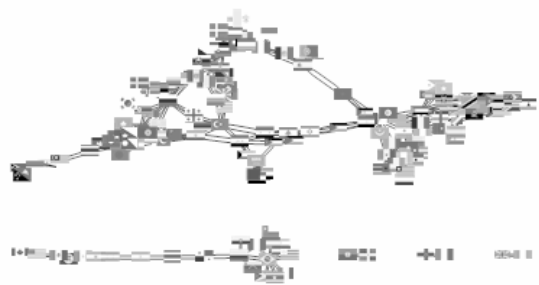

Fig. 4. Which countries share borders [19].

And what is important for our paper, it has a mode of operation where it can connect to social networks (e.g., Facebook) and reply to chat conversations automatically. In other words (as per our model) it is an information system on "other" transport protocol.

Aperator [22] makes tweets enable actionable commands on third-party web applications. Authors describe it as a new platform for application development. Since users can interact with third-party applications through the Twitter interface, Aperator demonstrates the possibility of purely back-end applications (Figure 5).

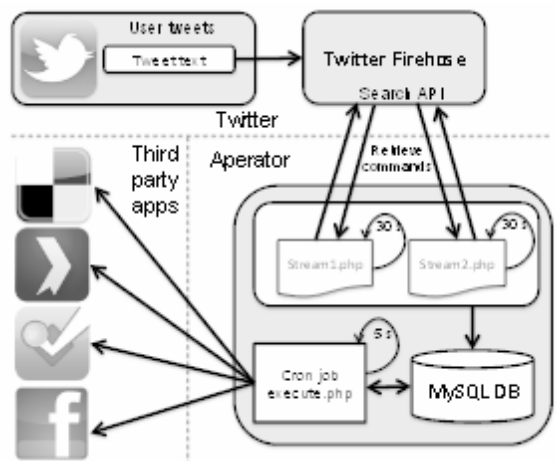

Fig. 5 Aperator architecture [22]

We can mention in this context M - Facebook's Human powered assistant [23]. It is built atop Facebook Messenger - the company's instant messaging application.

Of course, the classical example is IFTT [24]. IFTT lets connect different applications. It is based on the concept of recipes. Recipes are simple connections between products and apps. There are two types of Recipes: DO Recipes and IF Recipes. DO Recipes run with just a tap and enable users to create own personalized Button, Camera, and Notepad. IF Recipes run automatically in the background. They create powerful connections with one simple statement - if this then that. For instance, if some user uploads a file to his Dropbox folder, IFTTT might send a tweet or a text message or post a status update on any number of services, etc. The range of recipes permitted by IFTTT is extremely compelling.

The next example is API for bots in Telegram [25]. Bots are simply Telegram accounts operated by software – not people - and they'll often have Artificial Intelligence features. They can perform many operations - play, search, broadcast, remind, connect, integrate with other services, or even pass commands to the Internet of Things (IoT).

The idea for using messages (human or bots powered) with IoT applications is quite popular [26]. As one example, we can mention here Scouts project [27].

## IV. T411 FOR TWITTER

In this section, we present our service 411 for Twitter (T411). It is a platform for developing request-response services atop of Twitter. The name uses abbreviation from popular phone directory service in US – 411. Our first paper described this service (in Russian) has been published in INJOIT [28].

The main idea has been ported from our previous project for the SMS services platform. T411 lets turn any Twitter account into programmable auto-responder. It means that user-defined application will respond to incoming requests in Twitter.

For Twitter, incoming requests could be defined either as a mention or direct message. The examples below describe ant particular base account - @t411. For example, the typical mention looks like (it is a status in Twitter) so:

```
@t411  w msk
```

Actually, it is a real request. Any Twitter account, originated this status will get back a weekly weather prognosis for Moscow. And "get back" here means a mention or a direct message.

We assume that any message (mentions or direct messages) has got the following structure:

```
Key Optional_Text
```

So, in the above-mentioned example, *w* – is a key, *msk* is a text. It means that base account can host (provide) a set of services (each service depends on the own key). Keys are also user-defined. So, our prototype [29] lets users reserve a key for own service which will be based on @t411 account.

The next step in service creation is a web hook. It is a callback activated for a new message. Technically, it is a CGI script we can access via some URL. The result of its execution defines the response. On the top level the model for T411 service looks so:

```
Detect a new message => find a key =>
detect associated web hook => perform HTTP
request => create a response (tweet) with
results
```

So, any service in T411 is a pair <key, web hook>. The global architecture of T411 service is illustrated in Figure 6.

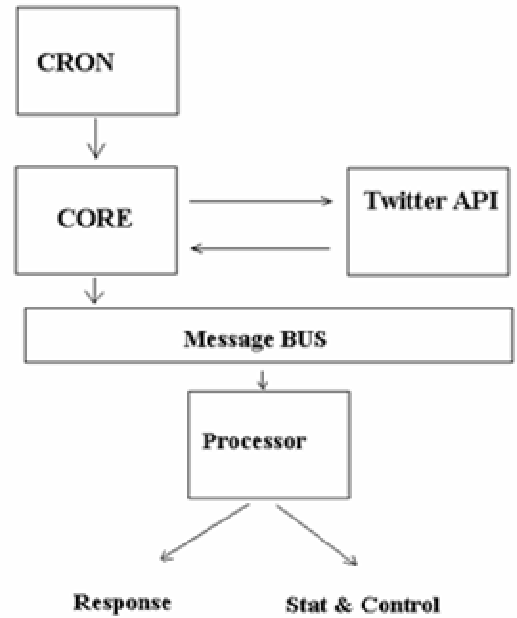

Fig. 6. T411 service [29]

In our prototype, Cron service has been implemented with Google AppEngine [30]. The core module has got own REST API, so we can start and stop it programmatically too. Cron module uses Core API and periodically starts its engine. This engine uses Twitter REST API for getting new messages (mentions and direct messages) for the particular account (it is @t411 in our example). Core Engine saves extracted messages in a dedicated Message Bus. The main System Processor accepts new entries from the Message Bus and evaluates them. The main data store is a typical key-value data base. A key here is a key for the service and a value is an appropriate web hook.

Any incoming message is a text. The processor extracts a key, find a web hook, perform HTTP request to CGI script and uses the response for a new tweet.

Technically, CGI script (web hook) could be hosted anywhere on the Internet. T411 always perform GET HTTP request. In this request service's processor passes a standard list of the parameters:

```
t – an original text
u – a name of Twitter's user (author of
a message)
```

Of course, this list could be extended.

The following JSP code illustrates stock information service. The registered key is t. So, all the messages starting with t will be processed by this service (bot). This bot expects messages in the following format:

`t stock_symbol`

for getting the quote. For example:

`t ORCL - quote for Oracle`

`t YNDX - quote for Yandex,   etc.`

As a web hook for this bot, we've set an URL for JSP file. E.g., in this particular case it is

`http://linkstore.ru/t411/quote.jsp`

So, for the incoming message

`@t411 t ORCL`

the processor will issue the following HTTP GET request:

`http://linkstore.ru/t411/quote.jsp?t=t%20ORCL`

And JSP file uses a couple of custom taglibs [31] for this task:

```
<%@ page contentType="text/plain; charset=utf-8" %>
<%@ taglib uri="taglib27.tld" prefix="get" %>
<%
String t = request.getParameter("t");
if (t==null) { out.println("unknown"); return; }
// the pattern is:
// t <space> stock_symbol
int i = t.indexOf(" ");
if (i<=0) { out.println(t+"?? could not get ticket"); return; }
t = t.substring(i+1).trim(); %>
<get:Quote symbol="<%=t.toUpperCase()%>" id="A" />
<%=A.get(0)+":    "+A.get(1)+" "+A.get(9)%>
```

Another standard example in T411 is the above-mentioned weather information service. It uses a key *w*, and text describes a city for weather info request. For example:

`@t411 w msk - weather in Moscow`

`@t411 w spb - weather in Saint-Petersburg`

Actually, the main model here is the mashup [32]. T411 lets either connect users to applications or connect different applications.

The whole *@t411* account presents itself a chatbot. Usually, chatbots are programmed by writing sets of pairs <text_pattern, text_effect>. The chatbot responds with the effect when the pattern matches the text received by the chatbot [33].

Patterns are simply regular expressions, and effects may include variables and conditionals (among more advanced structures).

So, you can send any question just to *@t411* account and get some dialogue with the system. As an analogue for this system we can mention the well-known Eliza system [34]:

`@abava> @t411 how are you?`

`@t411> @abava Let me see, just a minute please.`

## V. CONCLUSION

Our model service T411 presents a new way for Twitter deployment. It uses Twitter as a transport layer platform in information services and mashups. T411 demonstrates a new means of Twitter usage, which increases engagement for third party applications.

We can describe T411 as a new platform for application development. The users (the mobile users) can interact with third-party applications through the Twitter interface.

Also, T411 demonstrates a new way for monetization in Twitter. An ability to add application level interfaces to the standard account in Twitter could be a part of premium offerings for business-related accounts.

ACKNOWLEDGMENT

I would like to thank prof. Manfred Sneps-Sneppe from Ventspils University College for the valuable discussions.